\documentstyle[aps,prb,epsf,multicol]{revtex}
\tighten
\begin{document}
   \title{\Large \bf{
         Floquet states and persistent currents
         transitions in a mesoscopic ring
                    }
         }
\author{M. Moskalets$^{1}$ and
M. B\"uttiker$^2$
}
\address{
         $^1$Department of Metal and Semiconductor Physics,\\
        National Technical University "Kharkov Polytechnical Institute",
        Kharkov, Ukraine\\
        $^2$D\'epartement de Physique Th\'eorique, Universit\'e de Gen\`eve,
        CH-1211 Gen\`eve 4, Switzerland\\}

\date\today
   \maketitle
\bigskip

\begin{abstract}
We consider the effect of an oscillating potential
on the single-particle spectrum and the time-averaged persistent current of
a
one-dimensional phase-coherent mesoscopic ring
with a magnetic flux.
We show that in a ring with an even number of
spinless electrons
the oscillating potential has a strong effect on the persistent current
when the excited side bands are close to the eigen levels of a pure ring.
Resonant enhancement of side bands of the Floquet state generates
a sign change of the persistent current.

   \end{abstract}
   \ \\
   PACS: 72.10.-d, 73.23.-b, 73.23.Ra \\
\begin{multicols}{2}
\narrowtext

Mesoscopic systems subject to a periodic in time, external driving force
are now of considerable interest. This is illustrated by work concerning
low frequency ac-transport \cite{BTP94,PP94},
photon-assisted tunneling \cite{TG63,TE73,KJOENMS94,BHWKE95},
and quantum pumping \cite{SMCG99,Brouwer98}.  This list could easily be
extended.
Electrons interacting with a time-dependent potential
\cite{BL82}
can gain or loss energy and thus the electron system has no
stationary states and, in particular, there is no stationary ground state.
However if the external potential is periodic in time
we can describe the state of a system
using the Floquet function \cite{S65,W95,LR99}
which is a superposition of wave functions with energies
shifted by $n\hbar\omega$
(here $n$ is an integer; $\omega$ is the frequency of the
driving potential).
The existence of many components (side bands)
of a wave function has a strong effect
on the properties of a mesoscopic system.
For instance, side bands open up additional channels for transmission
through the mesoscopic system - photon assisted transmission
(see e.g., Refs. \onlinecite{W95,LR99}). The existence
of side bands is also a necessary condition
for pumping charge through
an unbiased mesoscopic sample \cite{MB02}.

The aim of the present paper is to investigate
the properties of a phase-coherent mesoscopic ring
in the Floquet state.
We are interested in the coherent properties of
the Floquet state
of a ring structures with an oscillating potential.
In a ring structure the wave functions must not only be
periodic in time but also periodic in space.
As a consequence a ring, in the presence of
an Aharonov-Bohm flux \cite{AB59}  $\Phi$ exhibits a persistent current
\cite{K70,BIL83,LDDB90,CWBKGK91,MCB93} .
The persistent current
is a signature of the coherence of the ground state of a mesoscopic ring
\cite{CB01}. Therefore, it is intersting to investigate how the
persistent current is affected by a time-dependent potential.
We find that under certain conditions the system
exhibits transitions
between the different components of the Floquet state.
Our analysis shows that the absolute value of the
amplitudes of the side bands of the Floquet
states are strong functions of frequency and of flux.
As a consequence the persistent current displays transitions
as a function of frequency and flux.

Let us consider the time-dependent Schr\"odinger equation for an electron
wave function $\Psi(x,t)$
on a circle of circumference $L$ threaded by
an Aharonov-Bohm magnetic flux $\Phi$
with an oscillating delta-function potential
\begin{equation}
\label{Eq1}
\begin{array}{l}
i\hbar\frac{\partial\Psi(x,t)}{\partial t} =
\hat H(x,t)\Psi(x,t), \\
\ \\
\hat H(x,t) = - \frac{\hbar^2}{2m}\left(
\frac{\partial}{\partial x}-2\pi i\frac{\Phi}{\Phi_0}
 \right)^2 + V(t)\delta(x), \\
\  \\
V(t) =  2LV\cos(\omega t). \\
\end{array}
\end{equation}

\noindent
Here $\Phi_0 = h/e$ is the magnetic flux quantum.
To solve this equation we will use the method of Floquet functions
\cite{W95,LR99}.
Because the Hamiltonian $\hat H$ depends on
time the system has no stationary eigenstates.
Since the Hamiltonian is periodic in time,
the states  of Eq.(\ref{Eq1})
can, according to the Floquet theorem,  be represented as a superposition of
wave functions with energies shifted by $n\hbar\omega$
\begin{equation}
\label{Eq2}
\Psi_E(x,t) = e^{-iEt/\hbar}\sum^\infty_{n=-\infty}\psi_n(x)
e^{-in\omega t}.
\end{equation}

By analogy with a pure ring problem we choose the functions
$\psi_n(x)$ in the following form

\begin{equation}
\label{Eq3}
\psi_n(x) = e^{2\pi i\frac{\Phi}{\Phi_0}\frac{x}{L}}
\left(a_n e^{ik_n x} + b_n e^{-ik_n x} \right).
\end{equation}
\noindent
Here $k_n = \sqrt{2mE_n/\hbar^2}$ and $E_n = E + n\hbar\omega$.
The coordinate $x$ is directed along the ring $0\leq x < L$.
Note, that for the evanescent modes ($E_n < 0$) we put
$k_n = i\kappa_n$ with $\kappa_n = \sqrt{2m|E_n|/\hbar^2}$.

On a ring the Floquet eigenfunction $\Psi_E$ must be periodic
in $x$. In addition its derivative is discontinuous at the
delta function barrier. Thus $\Psi_E$ is subject to the
following boundary conditions

\begin{equation}
\label{Eq4}
\begin{array}{l}
\Psi_E(x,t) = \Psi_E(x+L,t),  \\
\  \\
{\left.\frac{\partial\Psi_E(x,t)}{\partial x}\right|_{x=+0}} -
{\left.\frac{\partial\Psi_E(x,t)}{\partial x}\right|_{x=L-0}}   \\
\  \\
~~~~~~~~~~~~~~~~~~~~~~~ = \frac{2m}{\hbar^2}V(t)\Psi_E(0,t).
\end{array}
\end{equation}

\noindent These boundary conditions define the discrete set of Floquet
eigenenergies $E^{(l)}$ (where $l$ is an integer) and corresponding
Floquet eigenfunctions $\Psi_{E^{(l)}}$ which are characteristic
for the ring problem.
The quantization of the Floquet energy in a finite-size system
is quite analogous to
the quantization of an energy in the time-independent problem.
In addition each Floquet state can be occupied by only one electron
(because of the Pauli principle)
and thus the wave function $\Psi_{E}$ must be normalized

\begin{equation}
\label{Eq4a}
\begin{array}{l}
 \frac{1}{T}\int\limits_0^{T} dt
\int\limits_0^L dx |\Psi_E|^2
\equiv \sum\limits_n \int\limits_0^L dx |\psi_n|^2
= 1.
\end{array}
\end{equation}

\noindent Here $T = 2\pi/\omega$.
Furthermore, note that usually the Floquet energy $E$ is determined within
the
interval $0\leq E < \hbar\omega$.
However in our problem it is convenient not to reduce the discrete set of
$E^{(l)}$ to this interval.

Substituting Eq.(\ref{Eq2}) and Eq.(\ref{Eq3}) into the Eqs.(\ref{Eq4})
we obtain the following relations between coefficients $a_n$ and $b_n$
of the Floquet function $\Psi_E$ corresponding to the Floquet
energy $E$

\begin{equation}
\label{Eq5}
\begin{array}{l}
a_n A_n + b_n B_n = 0, \\
\  \\
a_n A_n - b_n B_n =  \\
\  \\
-i\frac{2\upsilon}{k_n}(a_{n-1} + b_{n-1} + a_{n+1} + b_{n+1}).
\end{array}
\end{equation}

\noindent Here we have introduced

\begin{equation}
\label{Eq5a}
\begin{array}{l}
A_n = e^{-2\pi i\frac{\Phi}{\Phi_0}} - e^{ik_nL}, \\
\ \\
B_n = e^{-2\pi i\frac{\Phi}{\Phi_0}} - e^{-ik_nL}, \\
\ \\
\upsilon = \frac{mLV}{\hbar^2}.
\end{array}
\end{equation}

Eqs.(\ref{Eq5}) couples amplitudes of different
index $n$. As a consequence we obtain
an infinite system of uniform linear equations for
the coefficients $a_n$ and $b_n$ ($n=0,\pm1,\pm2,\dots$).
To have a nontrivial solution
the corresponding (infinite range) determinant must be equal to zero.
This condition defines the allowed values of the Floquet energy
and the corresponding set of coefficients $a_n$ and $b_n$.

Using the method of continued fractions \cite{MR01}
the calculation of an infinite range determinant can be
greatly simplified.
To this end we rewrite the first equation of Eqs.(\ref{Eq5})
as follows

\begin{equation}
\label{Eq6}
b_n = -a_n \frac{A_n}{B_n},
\end{equation}

\noindent
Substituting the above relation into the second equation
of Eqs.(\ref{Eq5})
we obtain a recursive equation for the coefficients $a_n$.
It is convenient to introduce new quantities $x_n$ ($n\neq 0$)

\begin{equation}
\label{Eq7}
x_n = \frac{1}{\upsilon}\frac{a_n}{a_{n\mp 1}}
\frac{\sin(k_nL)}{\sin(k_{n\mp 1}L)}
\frac{B_{n\mp 1}}{B_n}.
\end{equation}

\noindent
Here and hereafter the upper (lower) sign is for $n\geq 1$ ($n\leq -1$).
In terms of the $x_n$ the recursive equation reads

\begin{equation}
\label{Eq8}
x_n = \frac{\sin(k_nL)}{k_nD_n -
\upsilon^2\sin(k_nL) x_{n\pm 1}},
\end{equation}

\noindent
where
\begin{equation}
\label{Eq9}
D_n = \cos\left(2\pi\frac{\Phi}{\Phi_0}\right) - \cos(k_nL).
\end{equation}

\noindent We can write the solution of Eq.(\ref{Eq8})
in the form of a continued fraction

\begin{equation}
\label{Eq10}
x_n = \frac{\sin(k_nL)}{k_nD_n -
\frac{\upsilon^2h_{n\pm 1}}{k_{n\pm 1}D_{n\pm 1} -
\frac{\upsilon^2h_{n\pm 2}}{k_{n\pm 2}D_{n\pm 2} -
\frac{\upsilon^2h_{n\pm 3}}{\ddots
}
}
}
}.
\end{equation}

\noindent Here
$
h_{n\pm 1} = \sin(k_nL)\sin(k_{n\pm 1}L),
$
\noindent
Using the quantities $x_n$ and Eq.(\ref{Eq7})
we can express any $a_n$ through the $a_0$

\begin{equation}
\label{Eq11}
a_n = a_0\upsilon^{|n|}\frac{B_n}{B_0}\frac{\sin(k_0L)}{\sin(k_nL)}
\prod_{j=\pm 1}^{n} x_i,~~~ n\neq 0.
\end{equation}

\noindent
The corresponding relation between $b_n$ and $b_0$ can be easily
obtained from the above equation and Eq.(\ref{Eq6}).

Now we can write down the equations containing only $a_0$ and $b_0$.
Using Eqs.(\ref{Eq5}) for $n=0$ and expressing $a_{\pm 1}$
and $b_{\pm 1}$ in terms of $a_0$ and $b_0$, respectively, we get

\begin{equation}
\label{Eq12}
\begin{array}{l}
a_0 A_0 + b_0 B_0 = 0, \\
\  \\
a_0 A_0 - b_0 B_0 =  \\
\  \\
-i\frac{2\upsilon^2}{k_0}(x_{-1} + x_{+1})(a_0 + b_0).
\end{array}
\end{equation}

\noindent
This system of equations has a nontrivial solution if its determinant
equals to zero
\begin{equation}
\label{Eq13}
\begin{array}{l}
k_0\left[ \cos\left(2\pi\frac{\Phi}{\Phi_0}\right) - \cos(k_0L) \right] \\
\ \\
- \upsilon^2(x_{-1} + x_{+1})\sin(k_0L) = 0 .
\end{array}
\end{equation}

\noindent
The solutions $k_0^{(l)}$ ($l = 0, \pm1, \pm2, \dots$)
of this (dispersion) equation give us a set of allowed Floquet
eigenenergies
$E^{(l)}$ = $(\hbar k_0^{(l)})^2$ $/(2m)$
and corresponding side bands
$E_n^{(l)}$ = $E^{(l)}$ + $n\hbar\omega$.

Note, that in the absence of an oscillating barrier ($\upsilon = 0$)
we obtain the well known spectrum of a perfect ring with a magnetic flux

\begin{equation}
\label{Eq14}
 E^{(l)}(\Phi) = \frac{h^2}{2mL^2}\left(l + \frac{\Phi}{\Phi_0}
\right)^2.
\end{equation}

\noindent For a weak potential ($\upsilon \to 0$) the Floquet energies
are close to those given by Eq.(\ref{Eq14}).

The main component
(corresponding to the energy $E^{(l)}$)
of the Floquet wave function
has a large amplitude: $a_0^{(l)}$ and/or $b_0^{(l)} \sim1$.
This is due to constructive interference in the ring.
In the general case, the amplitudes of side bands
(corresponding to energies $E^{(l)}_n$, $n\neq 0$)
are small due to destructive interference.
They are $a_n^{(l)}, b_n^{(l)} \sim\upsilon^{|n|}$
for a weak potential and
$a_n^{(l)}, b_n^{(l)} \sim\upsilon^{-|n|}$ for a strong potential.
Mathematically the effect of interference in a ring is described by
Eq.(\ref{Eq13}) (for the main component)
and by the denominator in Eq.(\ref{Eq8}) (for the side bands).
However there is a special (resonant) case when the amplitude of
a particular side band with an energy $E^{(l)}_{n_r}$
is comparable with the amplitudes of the main component
(corresponding to the energy $E^{(l)}$).
This is the case when the energy $E^{(l)}_{n_r}$ of the side band
is close to another Floquet eigenenergy $E^{(l')}$.
Note that at the same time the corresponding side band
$E^{(l')}_{-n_r}$ is close to $E^{(l)}$.

To determine the Floquet wave function $\Psi_{E^{(l)}}$
we need to know $a_0^{(l)}$
(then the coefficients
$b_0^{(l)}$, $a_n^{(l)}$ and $b_n^{(l)}$ can be found from
Eq.(\ref{Eq6}) and Eq.(\ref{Eq11})).
We find this coefficient from
the normalization condition Eq.(\ref{Eq4a}).
Substituting Eqs.(\ref{Eq2}), (\ref{Eq3}), (\ref{Eq6}), and Eq.(\ref{Eq11})
into Eq.(\ref{Eq4a}) we obtain

\begin{equation}
\label{Eq15}
|a_0^{(l)}|^2 = \frac{1}{L} \frac{\sin^2(k_0^{(l)}L/2 - \pi\Phi/\Phi_0)}{
Z^{(l)}\sin^2(k_0^{(l)}L)},
\end{equation}

\noindent where

\begin{equation}
\label{Eq16}
\begin{array}{l}
Z^{(l)} = \frac{\xi_0^{(l)}}{\sin^2(k_0^{(l)}L)} \\
\ \\
+ \sum\limits_{n\neq 0}\upsilon^{2|n|}
\frac{\xi_n^{(l)}}{\sin^2(k_n^{(l)}L)}
\prod\limits_{j=\pm 1}^{n}|x_j^{(l)}|^2,
\end{array}
\end{equation}

\noindent and

\begin{equation}
\label{Eq17}
\begin{array}{l}
\xi_n^{(l)} = 1 - \cos(k_n^{(l)}L)\cos\left(2\pi\frac{\Phi}{\Phi_0}
\right) \\
\ \\
+ \frac{\sin(k_n^{(l)}L)}{k_n^{(l)}L}\left[
\cos\left(2\pi\frac{\Phi}{\Phi_0} \right)
- \cos(k_n^{(l)}L)
\right].
\end{array}
\end{equation}

Next we consider the current carried by
the Floquet state $\Psi_{E^{(l)}}$. We will concentrate on
the time averaged (dc) current $I_{dc}$. To this end we integrate
the quantum mechanical current
\begin{equation}
\label{Eq17a}
I[\Psi] = i\frac{e\hbar}{2m}\left(
\Psi\frac{\partial \Psi^*}{\partial x} -
\Psi^*\frac{\partial \Psi}{\partial x}
\right) - \frac{e^2}{m}A\Psi\Psi^*,
\end{equation}
\noindent
(where $A = \Phi/L$ is a vector potential)
over the time period $T = 2\pi/\omega$
\begin{equation}
\label{Eq17b}
I_{dc}^{(l)} = \frac{1}{T}\int\limits_0^{T} dt I[\Psi_{E^{(l)}}(x,t)],
\end{equation}
\noindent
and obtain

\begin{equation}
\label{Eq18}
 I_{dc}^{(l)} = \frac{e\hbar}{m}
\sum\limits_{E_n^{(l)} > 0}
k_n^{(l)}\left(|a_n^{(l)}|^2 - |b_n^{(l)}|^2
 \right).
\end{equation}

\noindent
Note that the evanescent modes do not contribute to the current,
therefore we sum over the propagating
modes only ($E^{(l)}_n > 0$).

Now we can analyze the effect of an oscillating potential
on the electron wave function and the corresponding quantum mechanical
current. From the discussion given above, it follows that the oscillating
potential has in general a weak effect on the wave function which is mainly
determined by interference due to the ring geometry.
But as we already mentioned
this is not the case at resonance when the difference between
two Floquet eigenenergies is an integer number of
the energy quantum $\hbar\omega$.

The resonant frequencies are determined by the
level spacing. In the perfect ring investigated here
the spectrum has special features. Especially
at zero flux (or multiples of $\Phi_0/2$)
we have levels crossing each other.
Consequently for small flux there are two energy
scales which determine the resonant frequencies.
The first scale is the (large) level spacing
$\Delta^{(l)}$ = $E^{(l+1)}$ - $E^{(l)}$
that is a common feature of all finite-size systems.
For a small enough sample
$L\to 0$
the corresponding resonant frequency is large
$\omega_\Delta$ = $\Delta^{(l)}/\hbar$ $\sim L^{-2}$ $\to \infty$.
The second scale is determined by
the magnetic flux dependent separation
$\delta^{(l)}(\Phi)$ = $E^{(l)}(\Phi)$ - $E^{(-l)}(\Phi)$
between levels which are degenerate (in pairs) at zero flux
(or at multiples of $\Phi_0/2$) given by
$E^{(l)}(0)$ = $E^{(-l)}(0)$ (see Eq.(\ref{Eq14})).
This energy scale is a specific feature of ballistic,
perfect, ring-like structures.
The $n^{th}$ resonance $\delta^{(l)}(\Phi)$ = $n\hbar\omega$
occurs (in the case of a weak potential) at
\begin{equation}
\label{Eq19}
\omega = \frac{2\Delta^{(l)}}{n\hbar}\frac{\Phi}{\Phi_0}.
\end{equation}
\noindent
Note that at this condition the
$n^{th}$ side band $E^{(-l)}_{n}$ is in resonance with $E^{(l)}$
and vice versa: the
$-n^{th}$ side band $E^{(l)}_{-n}$ is in resonance with $E^{(-l)}$.

From Eq. (\ref{Eq19}) we can see that at small magnetic flux $\Phi\to 0$
the resonant frequency is small even for a small ring.
Thus we conclude that even a slowly (adiabatically)

\begin{equation}
\omega \ll \Delta/\hbar,
\label{Eq19a}
\end{equation}

\noindent
oscillating potential can essentially influence electronic
properties of a ring at small $\Phi \to 0$
(or at $\Phi\to\Phi_0/2$)
magnetic fluxes.
This regime is of interest in the present paper.
In what follows we describe numerical results
concerning the Floquet states in a ring with a small magnetic flux.

Let us consider a ring at zero magnetic flux
with a delta-function potential oscillating with a fixed frequency
$\omega$.
In this case the Floquet eigenenergies
$E^{(l)} = (\hbar k_0^{(l)})^2/(2m)$
(where $k_0^{(l)}$ is a solution of Eq.(\ref{Eq13}))
are the same as for a pure ring Eq.(\ref{Eq14}).
This is because the time averaged potential is zero.

Further, we choose some pair of degenerate (at $\Phi=0$)
Floquet energies $E^{(l)}$ and $E^{(-l)}$.
Because of interference the main component
$\psi_0$
of the Floquet wave function has a considerable amplitude:
$b_0^{(l)}\sim 1/L$,
$a_0^{(-l)}\sim 1/L$ (all other coefficients are close to zero).
Now we increase the magnetic flux which causes the levels
$E^{(l)}(\Phi)$ and $E^{(-l)}(\Phi)$
first to follow the energies
of a pure ring Eq.(\ref{Eq14}).
However close to the first resonance
$\Phi$ $\sim$ $\Phi_0\hbar\omega/(2\Delta^{(l)})$
they "interact" with the corresponding side bands
($E^{(-l)}_{+1}$ and $E^{(l)}_{-1}$, respectively)
and show an avoided crossing behavior.

 \begin{figure}
  \vspace{3mm}
  \centerline{
   \epsfxsize9cm
   \epsffile{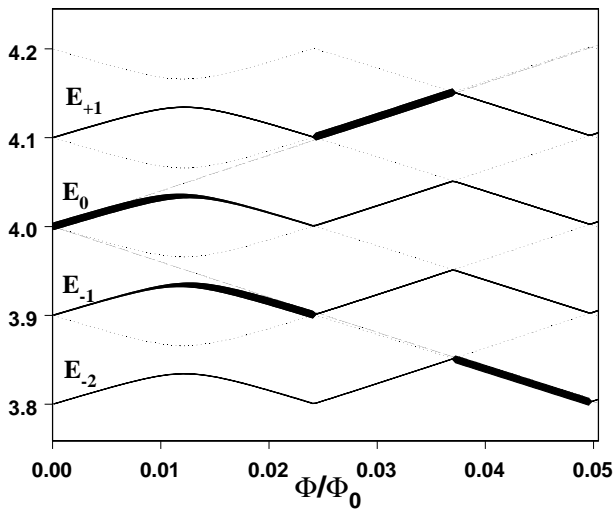}
             }
  \vspace{3mm}
  \nopagebreak
  \caption{
 The dependence of the Floquet eigenenergy
 $E_0$ = $E^{(l)}$ and some side bands $E_n$ = $E^{(l)} + n\hbar\omega$
 (solid lines; $n$ = $-2$, $-1$, $0$, $+1$) on the magnetic flux $\Phi$.
 The width of the solid lines is proportional to
 the probability of occupation of the corresponding side band.
 The energy and magnetic flux are given in units of
 $\epsilon_0$ = $4\pi^2\hbar^2/(2mL^2)$ and $\Phi_0 = h/e$, respectively.
 The parameters are:
 $l = 2$; $\hbar\omega = 0.1\epsilon_0$; $\upsilon = 0.05$.
 In addition the Floquet eigenenergy $E^{(-l)}$ with side bands
 ($n= -1; 0; +1; +2$) are also depicted by thin dotted lines.
 For comparison the eigenenergies
 $\epsilon_0\left(l \pm \Phi/\Phi_0 \right)^2$
 of a pure ring are depicted by thin dashed lines.
   }
\label{fig1}
\end{figure}

Close to resonance, because of constructive interference on the ring,
the amplitude of the wave function corresponding to the first side band
considerably increases \cite{note1}:
$a^{(l)}_{-1}\sim 1/L$ and $b^{(-l)}_{+1}\sim 1/L$.
The same occurs at the higher resonances.
Thus we can say that
at the resonance the Floquet state corresponds to an electron
equally distributed between two states with energies shifted by
$n\hbar\omega$ (for the $n^{th}$ resonance).
This is evident from Fig.\ref{fig1}
where we depict the dependence of the Floquet
energy $E^{(l)}$
and some side bands on the magnetic flux.
The width of the solid lines is proportional to
the square of the absolute value of the amplitude
( the probability of occupation ) of the side band
of the Floquet state.

We see that with increasing magnetic flux the particle
belonging to some Floquet eigenstate undergoes a transition
between states with an opposite direction of movement.
For instance, let us assume that at zero magnetic flux $\Phi = 0$
the particle is in state
$E_{0}^{(l)}$ with $a_{0}^{(l)} = 0$ and $b_{0}^{(l)} = 1/L$
and thus it carries a diamagnetic current
$I_{dc}^{(l)} < 0$ (see Eq.(\ref{Eq18})).
Then after the first resonance
(see Fig.\ref{fig1})
it passes into the state
$E_{-1}^{(l)}$ with $a_{-1}^{(l)} = 1/L$ and $b_{-1}^{(l)} = 0$.
In this case the particle carries a paramagnetic current $I_{dc}^{(l)} > 0$.
Correspondingly, after the second resonance
($\Phi$ $\sim$ $\Phi_0\hbar\omega/\Delta^{(l)}$)
the particle undergoes a transition into the state
$E_{+1}^{(l)}$ with $a_{+1}^{(l)} = 0$ and $b_{+1}^{(l)} = 1/L$
and it again carries a diamagnetic current, and so on.
Such a behavior has a strong effect on the persistent current
carried by the
(spinless) electrons on the ring.

Surprisingly, the pair of electrons occupying two Floquet sates
$E^{(l)}$ and $E^{(-l)}$ carry exactly the same current as
in the case of a pure ring:
\begin{equation}
\label{Eq20}
I_{dc}^{(l)} + I_{dc}^{(-l)} = -2I_0\frac{\Phi}{\Phi_0},
\end{equation}
\noindent
where $I_0 = eh/(mL^2)$.
Therefore the oscillating delta-function potential has no effect
on the persistent current in a ring with an odd number $N_{e}$
of (spinless) electrons
\begin{equation}
\label{Eq21}
I_{odd} = -N_{e}I_0\frac{\Phi}{\Phi_0},~~~~~|\Phi| < \Phi_0/2.
\end{equation}

However this is not the case for a ring with an even number of electrons.
In this case the current $I_{even}$
is mainly determined by the "unpaired" electron
in the highest occupied Floquet state.
As we discussed above the current carried by this electron
oscillates with a large amplitude.
In the low frequency limit (see Eq.(\ref{Eq19a})) of interest here
the period
$\delta\Phi$ $\sim$ $\Phi_0\hbar\omega/(2\Delta^{(l)})$
of these oscillations is much smaller
than the magnetic flux quantum $\Phi_0$.

\begin{figure}
 \vspace{3mm}
  \centerline{
   \epsfxsize9cm
   \epsffile{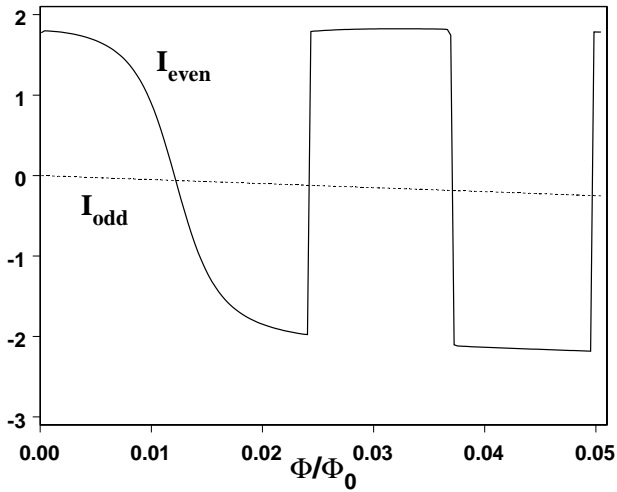}
             }
  \vspace{3mm}
  \nopagebreak
  \caption{
 Persistent current in a ring with an even $I_{even}$
 ($N_e = 4$; solid line)
 and an odd $I_{odd}$
 ($N_e = 5$; dashed line)
 number of electrons.
 The current is given in units of $I_0 = eh/(mL^2)$.
 The parameters are the same as in  Fig.1.
    }
\label{fig2}
\end{figure}

In Fig.\ref{fig2} we depict the dependence of the persistent current
on the magnetic flux
in a ring with four ($I_{even}$) and five ($I_{odd}$) electrons.

We would like to emphasize that the behavior of
the persistent current in a ring with an even number of electrons
is due to an interplay between the interference in a ring
and the excitation of side bands by an oscillating potential.
The persistent current reflects the behavior of a single Floquet state.
The interaction with an oscillating potential can not
lead to transitions between the different Floquet states.
As a consequence the particle stays in the same Floquet state
when the magnetic flux changes.
Because of interference the particle losses or gains some
energy quanta $\hbar\omega$
which brings it into the appropriate sub state of the
Floquet state when the magnetic flux goes through the resonant value.

However the interaction with an environment
can lead to transitions between different Floquet states.
In this case the particle will relax to the lowest unoccupied Floquet state
corresponding to a given value of a magnetic flux
and the peculiarities of $I_{even}$ will be diminished.
This effect will be considered elsewhere.

We remark on an essential difference between
the oscillations of the persistent current investigated here and
the oscillations of a persistent current with a period of $\Phi_0$.
In the ballistic case the energy is quadratic in
the magnetic flux Eq.(\ref{Eq14})
and the properties of a ring (in particular, the persistent current)
become periodic in $\Phi$ (with a period of $\Phi_0$)
only because of the relaxation to the state with a minimum energy
(for a more detailed discussion see  \cite{univ_ac}).
In contrast, the oscillations of interest here (see Fig.\ref{fig2} )
occur if only the system stays at the same Floquet state
(when the magnetic flux changes).

In the present paper we have considered a perfect ring.
But the effect under consideration is quite general
because it is due to a competition between
the quantum mechanical interference and the excitation of side bands by
an oscillating scatterer.
In particular, in the presence of disorder there is no level
degeneracy \cite{BIL83}.
However the oscillating scatterer can still generate
transitions between the different components of the Floquet state
(and thus can affect the persistent current)
if only an appropriate resonant condition is fulfilled:
$n\hbar\omega = E^{(l)}(\Phi) - E^{(l+1)}(\Phi)$
(here $E^{(l)}(\Phi)$ are energy levels in a ring with disorder).
The only difference from the perfect ring case is
that (in the low frequency limit)
the transitions are due to many photon processes: $n\gg 1$.

In conclusion, within the framework of the Floquet states approach
we have considered the effect of an oscillating delta-function
potential on the persistent current
in a ring of noninteracting spinless electrons
threaded by a magnetic flux.
We have found an unusual strong parity effect in a weak magnetic flux.
The current in a ring with an odd number of electrons
is diamagnetic and exactly the same as in a pure ring.
In contrast the current in a ring with an even number of electrons
oscillates in sign with a large amplitude and with a small
(compared with $\Phi_0$) period.

This work is supported by the Swiss National Science
Foundation.


\end{multicols}

\begin{thebibliography}{11}

\bibitem{BTP94}
   M. B\"{u}ttiker, H. Thomas, and A. Pr\^{e}tre,
   Z. Phys. B {\bf 94}, 133 (1994).

\bibitem{PP94}
   J. B. Pieper and J.C. Price, Phys. Rev. Lett. {\bf 72}, 3586 (1994).

\bibitem{TG63}
   P. K. Tien and J.P. Gordon, Phys. Rev. {\bf 129}, 647 (1963).

\bibitem{TE73}
   R. Tsu and L. Esaki, Appl. Phys. Lett. {\bf 22}, 562 (1973).

\bibitem{KJOENMS94}
   L. P. Kouwenhoven, S. Jauhar, J. Orenstein, P. L. McEuen, Y. Nagamune,
   J. Motohisa, and H. Sakaki, Phys. Rev. Lett. {\bf 73}, 3443 (1994).

\bibitem{BHWKE95}
   R. H. Blick, R. J. Haug, D. W. van der Weide, K. von Klitzing, and
   K. Eberl, Appl. Phys. Lett. {\bf 67}, 3924 (1995).

\bibitem{SMCG99}
    M. Switkes, C. M. Marcus, K. Campman, and A. C. Gossard,
    Science {\bf 283}, 1905 (1999).

\bibitem{Brouwer98}
    P. W. Brouwer, Phys. Rev. B {\bf 58}, R10135 (1998).

\bibitem{BL82}
   M. B\"{u}ttiker and R. Landauer, Phys. Rev. Lett. {\bf 49}, 1739 (1982).

\bibitem{S65}
   J. H. Shirley, Phys. Rev. {\bf 138 B}, 979 (1965).

\bibitem{W95}
  M. Wagner, Phys. Rev. A {\bf 51}, 798 (1995).

\bibitem{LR99}
   Wenjun Li and L. E. Reichl, Phys. Rev. B {\bf 60}, 15732 (1999).

\bibitem{MB02}
   M. Moskalets and M. B\"{u}ttiker, Phys. Rev. B {\bf  66}, 035306  (2002).

\bibitem{AB59}
  Y. Aharonov and D. Bohm, Phys. Rev. {\bf 115}, 484 (1959).

\bibitem{K70}
   I. O. Kulik, JETP Lett. {\bf 11}, 275 (1970).

\bibitem{BIL83}
   M. B\"{u}ttiker, Y. Imry, and R. Landauer,
     Phys. Lett. A {\bf 96}, 365 (1983).

\bibitem{LDDB90}
   L. P. Levy, G. Dolan, J. Dunsmuir, and H. Bouchiat,
   Phys. Rev. Lett. {\bf 64}, 2074 (1990).

\bibitem{CWBKGK91}
   V. Chandrasekhar, R.A. Webb, M.J. Brady, M.B. Ketchen, W.J. Gallagher
     and A. Kleinsasser, Phys. Rev. Lett. {\bf 67}, 3578 (1991).

\bibitem{MCB93}
   D. Mailly, C. Chapelier, A. Benoit,
     Phys. Rev. Lett. {\bf 70}, 2020 (1993).

\bibitem{CB01}
   P. Cedraschi and M. B\"{u}ttiker,
   Annals of Physics {\bf 289}, 1 (2001).

\bibitem{MR01}
   D. F. Martinez and L. E. Reichl,
   Phys. Rev. B {\bf 64}, 245315 (2001).

\bibitem{note1}
   When we consider the energy levels as a function of a magnetic flux
   it is convenient to use the classification of levels
   at zero magnetic flux.

\bibitem{univ_ac}
   M. V. Moskalets, Physica E {\bf 8}, 349 (2000).

\end{thebibliography}
 \end{document}